\documentclass[DIV16,twocolumn,10pt,headsepline,abstract=true]{scrartcl}
\usepackage[english]{babel}
\usepackage[utf8x]{inputenc}
\usepackage[intlimits]{amsmath}
\usepackage{graphicx}
\usepackage{float}
\usepackage{booktabs}
\usepackage{xspace}
\usepackage{url}
\usepackage{subfig}
\usepackage{color}
\usepackage[colorlinks,%
pagebackref=false, 
linktocpage=true, 
plainpages=false, 
bookmarksopen=true,%
bookmarksnumbered=true,%
pdfauthor={Pavel Hron},%
pdftitle={Application of reactive transport modelling to growth and transport of microorganism in capillary fringe},%
pdfsubject={},%
pdfkeywords={reactive modelling, capillary fringe, bacterial growth, adhesion, transport},%
]{hyperref}        
\usepackage[expproduct=cdot,alsoload={hep,binary}]{siunitx}
\usepackage{nicefrac}

\usepackage{ctable}
\usepackage{listings}

\addtokomafont{pagehead}{\small}
\addtokomafont{section}{\normalsize}
\addtokomafont{subsection}{\normalsize\normalfont}

\newunit{\years}{\text{year}}

\title{Application of reactive transport modelling to growth and
  transport of microorganisms in the capillary fringe}
\author{P.~Hron$^1$, D.~Jost$^2$, P.~Bastian$^1$, C.~Gallert$^3$, J.~Winter$^2$, O.~Ippisch$^{1,4}$}
\date{}
\publishers{\small \begin{flushleft}\noindent $^1$University of Heidelberg, Interdisciplinary Center for Scientific Computing, Im Neuenheimer Feld 368, 69120 Heidelberg, Germany\\\vspace{2mm}
$^2$Karlsruhe Institute of Technologie, Institut f\"{u}r Ingenieurbiologie und Biotechnologie des Abwassers, Am Fasanengarten Geb. 50.31, 76131~Karlsruhe, Germany \\\vspace{2mm}
$^3$University of Applied Science Hochschule Emden/Leer, Faculty of Technology, Microbiology – Biotechnology, Constantiaplatz 4, 26723 Emden, Germany \\\vspace{2mm}
$^4$Clausthal University of Technology, Department of  Mathematics, Erzstra{\ss}e 1, 38678 Clausthal-Zellerfeld, Germany \\\vspace{2mm}
  Email: pavel.hron@iwr.uni-heidelberg.de
\end{flushleft}
}

\newcommand{\Ecoli}{\textit{E.~coli}\xspace}
\newcommand{\DOC}{\mbox{$\mathrm{DOC}$\xspace}}
\newcommand{\Oxygen}{\mbox{$\mathrm{O_2}$\xspace}}
\newcommand{\Salpha}{\mbox{$s_{\alpha}\xspace$}}
\newcommand{\Sl}{\mbox{$s_l\xspace$}}

\newcommand{\Sg}{\mbox{$s_g\xspace$}}

\newcommand{\mumaxan}{\mbox{$\mu_{max,an}\xspace$}}
\newcommand{\mumaxa}{\mbox{$\mu_{max,a}\xspace$}}
\newcommand{\muan}{\mbox{$\mu_{an}\xspace$}}
\newcommand{\mua}{\mbox{$\mu_{a}\xspace$}}
\newcommand{\coxyw}{\mbox{$c_{l,O_2}\xspace$}}
\newcommand{\coxyg}{\mbox{$c_{g,O_2}\xspace$}}
\newcommand{\coxys}{\mbox{$c^*_{l,O_2}\xspace$}}

\newcommand{\cak}{\mbox{$c_{\alpha,\kappa}\xspace$}}
\newcommand{\cag}{\mbox{$c_{g,\kappa}^m\xspace$}}
\newcommand{\va}{\mbox{$\bf{v}_{\alpha}\xspace$}}
\newcommand{\vl}{\mbox{$\bf{v}_{l}\xspace$}}

\newcommand{\khcc}{\mbox{$k_H^{cc}\xspace$}}
\newcommand{\X}{\mbox{$c_{l,X}\xspace$}}
\newcommand{\Xp}{\mbox{$X_l\xspace$}}

\newcommand{\Xsmax}{\mbox{$c_{s,X}^{max}\xspace$}}
\newcommand{\Xs}{\mbox{$c_{s,X}\xspace$}}

\newcommand{\Xtotal}{\mbox{$c_{X}\xspace$}}
\newcommand{\Xptotal}{\mbox{$X_t\xspace$}}
\newcommand{\substrate}{\mbox{$c_{l,S}\xspace$}}

\newcommand{\Ysan}{\mbox{$Y_{S,an}\xspace$}}
\newcommand{\Ysa}{\mbox{$Y_{S,a}\xspace$}}
\newcommand{\Yo}{\mbox{$Y_{O_2}\xspace$}}

\newcommand{\Bo}{\mbox{$B_{O_2}\xspace$}}
\newcommand{\Bsan}{\mbox{$B_{S,an}\xspace$}}
\newcommand{\Bsa}{\mbox{$B_{S,a}\xspace$}}

\newcommand{\thetal}{\mbox{$\theta_l\xspace$}}

\newcommand{\thetas}{\mbox{$\theta_s\xspace$}}
\newcommand{\thetaalpha}{\mbox{$\theta_\alpha \xspace$}}

\newcommand{\Cells}{\mbox{$\text{cells}\xspace$}}

\newcommand{\rd}{\mbox{$d_c\xspace$}}
\newcommand{\rp}{\mbox{$p_d\xspace$}} 
\newcommand{\mo}{\mbox{$m_o\xspace$}}

\newcommand{\agw}{\mbox{$a_{gw}\xspace$}}
\newcommand{\adht}{\mbox{$a_{l,X}\xspace$}}
\newcommand{\katt}{\mbox{$k_{att}\xspace$}}
\newcommand{\kdet}{\mbox{$k_{det}\xspace$}}
\newcommand{\kpsi}{\mbox{$\psi\xspace$}}

\newcommand{\weight}{\mbox{$m\xspace$}}

\DeclareSIUnit\ecoli{\Ecoli{}}
\DeclareSIUnit\doc{\DOC}
\DeclareSIUnit\dw{dry weight}
\DeclareSIUnit\biomass{biomass}
\DeclareSIUnit\consumed{consumed}
\DeclareSIUnit\substr{substrate}
\DeclareSIUnit\glucose{glucose}
\DeclareSIUnit\oxygen{\Oxygen}
\DeclareSIUnit\cells{\Cells{}}
\DeclareSIUnit\pvolume{CF volume}
\DeclareSIUnit{\atmosphere }{atm}

\newcommand{\cdm}{\litre}
\newcommand{\ccm}{\milli \litre}
\newcommand{\temp}{$21\pm 1$\si{\celsius}}

\newcommand\T{\rule{0pt}{2.6ex}}       
\newcommand\B{\rule[-1.2ex]{0pt}{0pt}} 
\begin{document}

\maketitle

\begin{abstract}\noindent {
A multicomponent multiphase reactive transport simulator has been
developed to facilitate the investigation of a large variety of
phenomena in porous media including component transport, diffusion,
microbiological growth and decay, cell attachment and detachment and
phase exchange. The coupled problem is solved using operator
splitting. This approach allows a flexible adaptation of the solution
strategy to the concrete problem.

Moreover, the individual submodels were optimised to be able to
describe behaviour of \textit{Escherichia coli} (HB101 K12 pGLO) in
the capillary fringe in the presence or absence of dissolved organic
carbon and oxygen under steady-state and flow conditions. Steady-state
and flow through experiments in a Hele-Shaw cell, filled with quartz
sand, were conducted to study eutrophic bacterial growth and transport
in both saturated and unsaturated porous media. As \Ecoli{} cells can
form the green fluorescent protein (GFP), the cell densities,
calculated by evaluation of measured fluorescence intensities (in situ
detection) were compared with the cell densities computed by numerical
simulation. The comparison showed the laboratory experiments can be
well described by our mathematical model.
}
\end{abstract}

\section{Introduction}

There is a considerable ongoing effort aimed at understanding the
microbial growth and transport in porous media
\cite{Clement1997,Yarwood2006,Tufenkji2007,Chen2012}. Microbial
activity is of significant interest in various environmental
applications such as \textit{in situ} bioremediation \cite{Lee2009},
biodegradation of pollutants \cite{Bauer2008}, dispersal of
pathogenic microorganisms \cite{Unc2004}, protection of drinking
water supplies, and for subsurface geochemistry in general.

The design of bioremediation schemes requires understanding the
processes governing the growth, fate, cell attachment to solid
surfaces and transport of the microbes under the particular physical,
biological, and hydrogeological conditions involved
\cite{Ginn2002}. The appropriate processes can be influenced e.g. by
properties of the porous medium, the cells and pore water and gas
composition \cite{Mills1997, Or2007}.

The soil microbial population can be affected by the availability of
electron acceptors \cite{Sierra1995}, nutrients \cite{Reischke2013}
and bio-available water \cite{Skopp1990, Chang2003}. The capillary
fringe (CF), defined - in a wider sense than usual - as the region of
the subsurface above the groundwater table, but still dominated by
capillary rise, is a region where all these factors are abundantly
available and it offers attractive growth conditions for aerobic and
facultative anaerobic soil microorganisms
\cite{Affek1998,Jost2010,Jost2011}. It is thus a region where high
microbial activity is to be expected. In the presence of dissolved
molecular oxygen, most anaerobic metabolism is suppressed and aerobic
bacteria are the dominant groups in aerated soils. In regions heavily
contaminated with organic compounds the oxygen can be expected to be
quickly depleted, so that aerobic metabolism will no longer dominate
\cite{Kindred1989a}.

Reactive transport modelling is an essential tool for understanding
microbial growth and transport in the subsurface \cite{Steefel2005}
and has a significant impact e.g. on the treatment of contaminant
retardation, bioremediation or intrinsic biodegradation in the
subsurface, the description of nutrient fluxes and pollution
distribution and on the experimental design in general. Many studies
are often performed in laboratory scale experiments to identify the
principal processes that affect microbial transport and appropriately
designed mathematical models are developed and applied to predict the
retention and transport of microorganisms in natural conditions
\cite{Clement1997, Schaefer1998a,Schaefer1998}.

Many researchers have studied the behaviour of microorganisms in
porous media under saturated flowing conditions
\cite{Kouznetsov2004,Walker2004,Bradford2006},
but the transport in
unsaturated porous media is more complicated
\cite{Schafer1998,Powelson2001,Rockhold2007,Dechesne2008}. The
purpose of this study is to develop a mathematical model describing
multicomponent reactive transport processes in saturated and
unsaturated porous media (especially in the CF) at the Darcy scale, to
identify, quantify and investigate the relative importance of
processes on microbiological growth and transport of
\textit{Escherichia coli}, to characterise the processes by
appreciable microbial dynamics and to compare numerical simulations
with laboratory experiments.

\Ecoli{}, an indicator bacterium for faecal pollution of the
environment \cite{Chen2012}, can be used as a good example to
investigate bacterial growth and transport in partially saturated
porous media because \Ecoli{} cells may grow under aerobic and
anaerobic conditions \cite{Clark1989,Madigan2010}. In addition,
\Ecoli{} is easy to detect and quantify, and the net negative surface
charge and low inactivation rates ensure that they may travel long
distances in the subsurface \cite{Foppen2005}. These characteristics
make them a useful indicator of contamination of drinking water
supplies or as indicator bacterium of faecal pollution of the
environment \cite{Chen2012} and is therefore a frequent research
objective of many scientists
\cite{Lacoursiere1986,Duffy1999,Powelson2001,Jiang2007}.

\section{Model development}

We assume that an unsaturated, incompressible porous medium under
investigation consists of three phases $\mathcal{P}=\{s,l,g\}$: solid
phase $s$ and two fluids $l,g$, liquid and gas, respectively as in
\cite{Allen1985}. Each of these phases $\alpha \in \mathcal{P}$ is
composed of a set of components~$\mathcal{K}_\alpha$.

Each phase $\alpha$ has its own mass phase density $\rho_\alpha$ and
volume fraction $\thetaalpha$. From their definitions, the volume
fractions clearly must fulfil the constraint $\sum_\alpha
\thetaalpha=1$. The relation between porosity $\phi$, aqueous and
gaseous saturation $\Salpha$ and its content is $\thetaalpha=\phi
\Salpha$, where $\Sl+\Sg=1$.

For each component $\kappa \in \mathcal{K}_\alpha$, the general
macroscopic mass conservative equations describing the transport and
reaction of aqueous and solid phase species are written as
\begin{equation} \frac{\partial (\thetaalpha \cak)}{\partial t} +
\nabla\cdot\{\cak \va + j_{\alpha,\kappa} \} =
R_{\alpha,\kappa}, \label{Eq:FullModelMoleBalance}
\end{equation} where $\cak$ is the mass concentration of component
$\kappa$, $\va$ is the velocity of phase $\alpha$, $j_{\alpha,\kappa}$
is the diffusion term and $R_{\alpha,\kappa}$ describes the
reaction.

We assume as in \cite{ABT} and \cite{Class2002} that sum of the
diffusive fluxes for all components in each phase is zero
\begin{equation}
  \label{Eq:DiffusivityConstraint} \sum_{\kappa\in\mathcal{K}_\alpha}
j_{\alpha,\kappa} = 0,
\end{equation} which ensures that the phase $\alpha$ moves with the
phase velocity $\va$. In a typical immobile species transport equation
\eqref{Eq:FullModelMoleBalance}, the advection and diffusion are zero
and only the reaction term exists.

The phase velocity is related to the phase pressure $p_\alpha$ via the
extended Darcy law
\[
\va = -\frac{k_{r\alpha}(s_\alpha)}{\mu_\alpha} K
\left(\nabla p_\alpha - \rho_\alpha g\right),
\]
where $k_{r\alpha}$ is the nonlinear relative
permeability function, $K$ is the scalar absolute permeability (in
general it can be tensor, \cite{Helmig1997}), $\mu_\alpha$ is the
dynamic viscosity of the fluid and $g$ is the gravitational
acceleration.

The model \eqref{Eq:FullModelMoleBalance} is completed by the
macroscopic capillary pressure-saturation relationship
\begin{equation}\label{Eq:CapillaryPressure} p_g - p_l = p_c(s_l).
\end{equation} Various static capillary pressure-saturation models by
\cite{Brooks1964} or van \cite{Genuchten1980} were developed based on
laboratory experiments.

The Ficks' law is used together with the second model of Millington
and Quirk for the dependence of the effective diffusion coefficient on
phase saturation \cite{Jin1996} and one obtains
\[ j_{\alpha,\kappa} = - s_{\alpha}^2 \phi^\frac{4}{3}
D_{\alpha,\kappa} \nabla \cak,
\] where $D_{\alpha,\kappa}$ is the molecular diffusion
coefficient of component $\kappa$ in phase $\alpha$.

\subsection{Operator splitting approach}

Many simulators for reactive multiphase flow see
e.g. \cite{Class2002,MayerFrindBlowes,Bielinsky2006,Podgorney_2012b} use
a global implicit approach with full upwinding of convective
terms. This results in a large amount of numerical diffusion for the
component transport with a subsequent overestimation of reaction.  We
follow \cite{Clement1998,XuPruess2001} and \cite{Toughreact2006} and
choose an operator splitting approach.

In a first step the system \eqref{Eq:FullModelMoleBalance} is split
into a transport and into a reaction part
\begin{subequations}
\begin{align} \frac{\partial (\thetaalpha \cak)}{\partial t} +
\nabla\cdot\{\cak \va + j_{\alpha,\kappa} \} &=
0, \label{Eq:FullModelTransportReactiona}\\ \frac{\partial
(\thetaalpha \cak)}{\partial t}&=
R_{\alpha,\kappa},\label{Eq:FullModelTransportReactionb}
\end{align}
\label{Eq:FullModelTransportReaction}
\end{subequations} then the transport part is split again in a phase
transport and a phase composition part.

\subsubsection*{Phase transport}

Summation of the transport part \eqref{Eq:FullModelTransportReactiona}
over all components of each phase and using assumption
\eqref{Eq:DiffusivityConstraint} yields a mass balance equation for
each liquid phase
\begin{equation}\label{Eq:PhaseBalance} \frac{\partial \left(
\thetaalpha \rho_\alpha \right)}{\partial t} + \nabla\cdot\{
\rho_\alpha \va \} = 0.
\end{equation} The liquid mass phase density remains constant, while
$\rho_g$ may depend on phase composition. Both phases are assumed to
be mobile, which is needed to model active gas transport, entrapment
of gaseous species or gas production by microorganisms.

\subsubsection*{Component transport}

Selecting a \textit{reference} component $\kappa_{\alpha,0} \in
\mathcal{K}_\alpha$ in each liquid phase, e.g. water in the liquid
phase and air in the gas phase, it remains to solve the
$|\mathcal{K}_l| + |\mathcal{K}_g|-2$ component balance transport
equations in the form
\begin{equation}
  \label{Eq:ModelComponentBalance} \frac{\partial (\thetaalpha
\cak)}{\partial t} + \nabla\cdot\{\cak \va + j_{\alpha,\kappa} \} = 0.
\end{equation}

\subsubsection*{Reaction}

Aqueous species are subject to local chemical interactions with the
solid and gaseous phase by \eqref{Eq:FullModelTransportReactionb} or
are assumed to be at local equilibrium. The term $R_{\alpha,\kappa}$
includes all possible kind of reactions and it can be divided into a
chemical or biological reaction part $r_{\alpha,\kappa}$, phase
exchange between fluids $e_{\alpha,\kappa}$ and interaction between
the liquid and solid phases $a_{\alpha,\kappa}$. Direct interaction
between solid and gas phase is not considered.

\subsection{Primary variables}

The model \eqref{Eq:FullModelMoleBalance} split into
(\ref{Eq:FullModelTransportReaction}-\ref{Eq:ModelComponentBalance})
is a mixture of partial differential equations (PDEs), ordinary
differential equations (ODEs) and algebraic equations. The PDEs and
ODEs can be solved using different choices of primary variables and
the remaining unknowns can be computed by solving algebraic equations.

In our applications we use the liquid phase and capillary pressures
$p_l, p_c$ and the concentrations $c_{\alpha,\kappa}$ of all but one
component $\kappa\in\mathcal{K}_\alpha\setminus\{\kappa_{\alpha,0}\}$
for each fluid phase as primary variables. Concentrations in solid
phase are always taken as primary variables.

Given these primary variables the remaining quantities can be computed
as follows: water saturation is determined via the inverse function
$p_c^{-1}$ of \eqref{Eq:CapillaryPressure}. Because $\rho_l$ is
constant, the remaining concentration is given by the
$|\mathcal{K}_l|-1$ known concentrations.

The gas molar density $\nu_g$ is related to gas mass density by
$\rho_g=\nu_g M_g$, where $M_g$ is the average molar mass of gas
mixture. Total pressure in the gas phase is related to the sum of the
partial pressures (Dalton's law)
\[\sum_{\kappa\in\mathcal{K}_g} p_{g,\kappa} = p_g. \] By means of the
ideal gas law, partial pressures relate to the molar concentrations
$\cag$ of gas species
\[p_{g,\kappa}=RT \cag\] with $R$ being the ideal gas law constant and
$T$ the temperature of the system as in \cite{Molins2007}. In our
applications, we assume constant temperature~$T$ across the considered
domain.

\subsection{Solution procedure}

The numerical solution of two-phase flows \eqref{Eq:PhaseBalance}
employs space discretization on a structured grid using a cell-centred
finite volume method (CCFV) with two-point flux approximation and full
upwinding in capillary pressure \cite{Neumann2013}. Time is
discretized fully implicitly to achieve unconditional stability. The
arising nonlinear equations are linearised with Newton-Method, the
Jacobi matrix is derived through numerical differentiation and the
resulting linear system is solved with BiCGStab iterative solver with
an algebraic multigrid preconditioner \cite{Blatt2010}.

To solve the solute transport \eqref{Eq:ModelComponentBalance} in
advection dominated cases, the CCFV method using second order Godunov
reconstruction of upwind fluxes (\textit{minmod} slope limiter)
\cite{LeVeque1990,LeVeque2002} together with the explicit Euler
scheme in time is used. This scheme for advective transport gives
accurate solution and reduces numerical diffusion. For the solution of
\eqref{Eq:ModelComponentBalance} in diffusion dominated cases (mainly
transport of gaseous species), we combine the CCFV method with central
differencing scheme and implicit Euler time step discretization.

The system of ODEs \eqref{Eq:FullModelTransportReactionb} was solved
element wise using embedded Runge-Kutta-Fehlberg~$4(5)$ method
\cite{Hairer1993} which allows automatically determined adaptive step
size to reduce numerical error.

All the numerical methods were implemented and the numerical
simulations were performed in the {DUNE} simulation framework
\cite{Dune2008a, Dune2008b, pdelabalgoritmy}. The numerical code is
parallelized using domain decomposition and can be applied to simulate
two- and three-dimensional systems containing relatively high number
of degree of freedoms.

A sequential non-iterative approach (SNIA) similar to
\cite{XuPruess2001} was used. After solution of the flow equations
\eqref{Eq:PhaseBalance}, the fluid velocities and phase saturations
are used for component transport simulation. The transport model
\eqref{Eq:ModelComponentBalance} including advection and diffusion is
solved on a component basis. The resulting cell concentrations
obtained from the transport are substituted into the chemical reaction
model \eqref{Eq:FullModelTransportReactionb}.
In many studies on reactive flow, a sequential iteration approach
(SIA) was used, e.g. \cite{Yeh1989,Simunek1994}. For sufficiently
small Courant-Friedrichs-Lewy number ($CFL$) the difference between
SNIA and SIA is generally small (especially involving reactions with
slow kinetic rates) and an operator splitting error is reduced
significantly \cite{Xu1999}. Because the explicit time discretization
for component transport is used, the CFL condition is a necessary
condition for stability and the transport time step in applications is
restricted to fulfil $CFL=0.4$.

If the advection of all fluid components is negligible (diffusive
processes dominate advection) or if the flow is stationary and there
are no large concentration gradients in the domain, the system of
transport and reaction equations \eqref{Eq:FullModelTransportReaction}
can be coupled again and solved fully implicitly without operator
splitting. This method is called direct substitution approach (DSA)
and was discussed in
e.g. \cite{Steefel1994,Saaltink1998,Saaltink2004}.

In the overall scheme the time step for two-phase system is restricted
only with convergence of Newton-Method (arising system is highly
nonlinear) and may be larger than the time step for the transport
equation. Then several steps for the component transport/reaction
system are applied with one two-phase time step. If DSA approach is
used, the time step size is dependent on the convergence of non-linear
solver \cite{Saaltink2001}.

To reduce operator splitting error between phase transport and
component transport, the maximal change in a cell saturation in one
time step is limited to $0.1$.

\section{Microbial processes and model development}

Aside from hydrogeological and physical conditions, bacterial
transport and growth in the subsurface is influenced by many
processes \cite{Ginn2002} . The nature of these processes
represents a challenge for reactive transport modelling in that one
biological mechanism is often dependent on and/or influenced by
another mechanism. Thus, it may be necessary to consider the
interdependency of the various processes, which are relevant for
microbiological growth in unsaturated and saturated porous media under
steady-state and transient conditions.

We will introduce basic models based on a macroscopic approach for
modelling the transport of microorganisms in saturated and unsaturated
porous media, which were motivated, guided and optimised based on
laboratory batch and flow-through experiments and measured data for
\Ecoli{}, strain HB101 K12 pGLO.

\subsection{Modelling attachment processes in porous media}

In many laboratory and field studies, models based on a colloid
filtration theory (CFT) have been used extensively to evaluate
microbial transport and adhesion in saturated and unsaturated porous
media, e.g. \cite{Ginn2002,Tufenkji2007}.

Assuming that the attachment of \Ecoli{} cells to sand grains is
reversible, the general formulation of the adhesion term $\adht$
describing cell adhesion from liquid phase is written as
\cite{Bradford2006}
\begin{equation} \adht = -\thetal \katt \kpsi \X + \thetas \kdet
\Xs, \label{Eq:adhesionliquid}
\end{equation} where $\X$ is the \Ecoli{} concentration in the liquid
phase, $\Xs$ is the \Ecoli{} concentration in the solid phase, $\katt$
is the first-order deposition coefficient, $\kdet$ is the first-order
detachment coefficient and $\kpsi$ is a dimensionless deposition
function. The solid phase mass balance equation for \Ecoli{} is given
as
\begin{equation} \frac{\partial \left(\thetas \Xs \right)}{\partial
t}=-\adht. \label{Eq:adhesionsolid}
\end{equation}

To account for time dependent deposition behaviour and decreases in
the solid surface area available for bacterial adhesion, a general
form for $\kpsi$ is utilised as \cite{Schafer1998}
\begin{equation} \kpsi = \left(1-\frac{\Xs}{\Sl
\Xsmax}\right), \label{Eq:deposition}
\end{equation} where $\Xsmax$ is the maximum attainable bacterial
concentration on the solid surface. The solid-liquid interstitial
available for cell attachment decreases with water content. As $\Xs$
rises, the bacteria render the solid surface less attractive for
further attachment and the deposition function \eqref{Eq:deposition}
will decrease to zero. The maximum concentration of $\Xs$ will be
equal to $\Sl \Xsmax$.

\subsection{Modelling microbial growth kinetics}

Different batch-culture experiments under varying conditions showed
\cite{Hron2014}, that the kinetics of anaerobic and aerobic
growth of \Ecoli{} is limited by the availability of dissolved oxygen
$\coxyw$ and bioconvertible dissolved organic carbon (\DOC{}), denoted
by $\substrate$. It was observed, that anaerobic growth takes place
only if the amount of available oxygen is very low. If enough oxygen
is available only aerobic respiration is active and the growth of
\textit{E.~coli} is faster.

Furthermore, experiments with different oxygen concentrations showed
the dependency of biomass production on oxygen availability. With
decreasing oxygen concentration the total biomass production also
declined. For constant air oxygen concentration $1\text{-}4\%$, the
$\DOC$ was depleted even slower than in the anaerobic case. It can be
explained by switching between aerobic respiration and anaerobic
fermentation, which may take an additional time.

Because the cells attached to the sand grains are in direct contact
with the liquid phase, we can define a \textit{total} liquid \Ecoli{}
concentration as $\Xtotal = \X + \Xs \thetas \slash \thetal$. We
assume, that cells in both phases are able to consume nutrients and to
starve, but new cells become immediately mobile in the liquid phase
and are able to attach only through attachment processes. This
assumption is motivated by results of flow-through experiments with
continuous $\DOC$ supply, where the measured cell concentration in the
outflow was high (more then $\SI{1.e8}{\cells \per \ccm} \text{
water}$) and the typical assumption for subsurface porous media that
the majority of cells are attached to the solid is not valid.

The anaerobic and aerobic specific growth rates based on the Contois
model \cite{Contois1959} can be expressed as
\begin{subequations}
\begin{align}
\mua &= \mumaxa
\frac{\substrate}{\Xtotal \Bsa + \substrate} \frac{\coxyw}{\Xtotal \Bo
+ \coxyw}, \label{Eq:double-specific-growth} \\
\muan &= \max(\mumaxan \frac{\substrate}{\Xtotal \Bsan +
\substrate}-\mua,0), \label{Eq:specific-growth}
\end{align}
\label{Eq:contoismodel}
\end{subequations}
where $\mumaxa$ and $\mumaxan$ are aerobic and anaerobic specific
growth rates, respectively The constants $\Bsa,\Bsan$ and $\Bo$
multiplied by $\Xtotal$ correspond to the half-saturation constants in
a well known Monod kinetics \cite{Monod1949}. The Contois
modification is able to overcome better the difficulty described in
\cite{Kovarova-Kovar1998}, where the authors mentioned that the
half-saturation constant in Monod model could vary even during a
single growth cycle. To simulate bacterial growth in subsurface under
different growth conditions, a single set of parameters, which is able
to describe all possible states, is needed.

If the dissolved oxygen concentration is low, the cells are only
growing anaerobically. With increasing oxygen concentration, they are
growing in both modes but not faster than in situation when sufficient
dissolved oxygen is available.

The reaction terms denoting growth and decay (with a rate constant $\rd$) of cells
in the liquid and solid phase are given by
\begin{subequations}
\begin{align} r_{l,X}&=\thetal \left( \left(\mua + \muan\right) \Xtotal - \rd \X \right), \\
r_{l,S}&= -\thetas \rd \Xs.
\end{align}
\label{Eq:growth}
\end{subequations} The balance equation for the consumable substrate
and dissolved oxygen have the form
\begin{subequations}
\begin{align} r_{l,S}&= - \thetal \left( \frac{\mua}{\Ysa} +
\frac{\muan }{\Ysan} \right) \Xtotal, \\ r_{l,O_2} &= - \thetal \left(
\frac{\mua} {\Yo} + \mo \coxyw\right) \Xtotal,
\end{align}
\label{Eq:substratebalance}
\end{subequations} where the yield coefficients $\Ysa,\Ysan$ and $\Yo$
are the links between growth and substrate utilisation and $\mo$ is the non-growth-associated maintenance factor for oxygen. The
maintenance rate is also dependent on oxygen availability.

\subsection{Phase exchange in porous media}

Even for the growth experiments in batch-culture with continuous
shaking and instantaneous oxygen supply the measurements showed a
lower oxygen content in the liquid phase that in equilibrium with the
gas phase. The consumption of oxygen in water thus was faster than
oxygen dissolution in air and the local equilibrium can not be
assumed.

To specify macro-scale closure relationships for the mass exchange of
oxygen in porous media between the gas and the water phase, we follow
the model discussed in \cite{Geistlinger2005} and \cite{Holocher2003}
based on the stagnant film model for spherical gas bubbles. The
interphase mass flux is proportional to the concentration deficit in
the water phase \cite{Mayer1996} and can be described by
\begin{equation} - e_{g,O_2} = e_{l,O_2}=\beta \agw \left(\coxys -
\coxyw \right), \label{Eq:massTransfer}
\end{equation} where $\beta$ is the mass exchange coefficient, $\agw$
is the effective gas-water interface and $\coxys$ is the equilibrium
oxygen concentration in the liquid phase. For gases with a low
solubility, like oxygen in water, the local equilibrium concentration
of oxygen in water $\coxys$ can be described by a Henry's law
\[ \coxys= \khcc \coxyg
\] with constant $\khcc$ and oxygen concentration in air
$\coxyg$, see \cite{Sander1999}.

\subsubsection{Mass exchange coefficient}

The mass exchange coefficient $\beta$ for a spherical structure with
the harmonic mean particle diameter $\rp$ is given by \cite{Clift1978}
\begin{equation} \beta
=D_{l,O_2}\left(\frac{2}{\rp}+\frac{1}{\delta}\right),\label{Eq:masstransfercoeff}
\end{equation} where $D_{l,O_2}$ is the oxygen diffusion coefficient
in water and $\delta$ is the thickness of the stagnant film layer.

For advection flow regimes, the boundary layer thickness depends on
the interface velocity. The difference between flow and
interface velocity directly at the interface is in most cases very
small and therefore neglected \cite{Niessner2011}. In this case, the
thickness of the film layer can be expressed by
\[ \delta=\sqrt{\frac{\pi \rp D_{l,O_2}}{\|\vl\|}}.
\]
Comparison of different mass exchange
coefficient-velocity correlations for oxygen can be found in
\cite{Geistlinger2005}.

The gas-liquid interface plays a crucial role in the phase exchange in
unsaturated porous media. Many relationships used to predict
gas-liquid specific interfacial area, derived either from pore-scale
network models, from Lattice-Boltzman simulations or from experiments
can be found in literature,
see e.g. \cite{Cary1994,Ahrenholz2011,Brusseau1997, Chen2006, Kim1999,
Schaefer2000, Porter2010}.

A basic model - based on geometrical relationships - to estimate the
total surface area for any packing of spheres was presented by
\cite{Gvirtzman1991}. The total effective interstitial surface area is
given by
\begin{equation} \agw = \kappa \Sg
\frac{6\phi}{\rp}, \label{Eq:surfaceAreaBasic}
\end{equation} where $\kappa$ is the fraction of air bubble surface
area exposed to mobile water \cite{Geistlinger2005}.

The gas-liquid interface in \eqref{Eq:surfaceAreaBasic} is
proportional to the gas content. However, the pore network models
\cite{Reeves1996, Joekar-Niasar2007} and experiments with glass beads
\cite{Culligan2004, Porter2010} proved this relationship only for
$\Sl>0.3$ and the model \eqref{Eq:surfaceAreaBasic} may overestimate
the gas-liquid interface and thus phase exchange in regions with low
water saturation.

The model \eqref{Eq:massTransfer} together with mass exchange
coefficient \eqref{Eq:masstransfercoeff} and interstitial surface area
\eqref{Eq:surfaceAreaBasic} describes a first-order kinetic, where the
exchange rate depends on water content and only one free parameter
$\rp$. The sensitivity of the model regarding to $\rp$ is discussed in
numerical results.

\section{Laboratory experiments and numerical simulations in porous
  media}
\subsection{Porous medium}

In experiments to quantify growth and attachment of \Ecoli{}, quartz
sand (diameter: $0.2-\SI{0.6}{\milli \meter}$, porosity $0.39$) was used
as a model porous medium. A multistep outflow experiment (MSO) was
performed to investigate the flow-rate dependence of unsaturated
hydraulic properties of the sand and hydraulic conductivity together
with parameters for the van Genuchten model were
estimated, see Table \ref{Tab:hydrparam}.

\begin{table}[h]
 \centering
 \caption{Hydraulic parameters}
 \smallskip
  \small\addtolength{\tabcolsep}{5pt}
 \label{Tab:hydrparam}
 \begin{tabular}{l l }
  \toprule
  \multicolumn{2}{l}{van Genuchten model} \T\B\\ \hline
   $n$ & 5.48 \T \\
   $\alpha$ & \SI{1.21e-3}{\per \pascal}
   \B \\ \midrule
   porosity $\phi$& $0.39$ \T \\
   permeability  $K$& $\SI{2.6E-011
}{\meter \squared}$ \T \\ \bottomrule
   \end{tabular}
\end{table}

\subsection{Adhesion parameters}

In the work by \cite{Lutterodt2012} many flow through experiments were
conducted to investigate adhesion kinetics of different \Ecoli{}
strains. The adhesive rates for different strains may differ by many
orders of magnitude ($\num{e-2}\text{-}\SI{e-7}{\per \second}$) and
the adhesive kinetic needs to be established for every strain
separately. For this reason we conducted similar flow through
experiments to estimate the parameters in model
\eqref{Eq:adhesionliquid} for strain HB101~K12~pGLO.  The flow through
experiments were conducted in stainless steel capillaries (diameter
\SI{0.4}{\centi \meter}, length \SI{10}{\centi \meter}, filled with
quartz sand) at initial bacterial concentrations ranging from
\num{0.7e8} to \num{1.0e9} highly active \si{\cells \per \ccm} and
pore velocities between \num{1.3} and \SI{4.}{\meter \per \day}. The
number of cells remaining in \SI{0.9}{\percent} $\mathrm{NaCl}$
suspension was counted in the outflow, and flow through curves were
determined.

The resulting data were then used for parameter estimation. The
inverse modelling was done with a computer model which uses the
Levenberg-Marquardt-Algorithm for the parameter estimation with
sensitivities derived by numerical differentiation. The forward model
solves the transport equation for \Ecoli{} concentration in water
\eqref{Eq:ModelComponentBalance} together with adhesion kinetic models
\eqref{Eq:adhesionliquid} and \eqref{Eq:adhesionsolid} with given
initial concentrations. To eliminate numerical discretization errors
in solving forward problem, a one-dimensional domain was divided into
$512$ elements for CCFV simulation. In each time step the
concentration in the last element was taken as outflow
concentration. The attachment coefficients in \cite{Bradford2006} have
values between $\num{7.7e-3}$ and $\SI{1.3e-4}{\per \second}$, which
is in a good agreement with our estimated attachment coefficient
$\katt = \SI{3.e-4}{\per \second}$. The detachment coefficient
$\kdet=\SI{6.2e-6}{\per \second}$ is also comparable with values
$\num{1.7e-8}\text{-}\SI{3.3e-6}{\per \second}$ reported in
\cite{Bradford2006} with pore velocity about $\SI{4}{\meter \per
\day}$. The estimated maximum attainable concentration $\Xsmax$ was
$\SI{1.6e8}{\cells \per \ccm}\text{ porous
  medium}$. \cite{Schafer1998} measured the value of approx. $\SI{0.5e8}{\cells \per \ccm}\text{ porous
  medium}$ for \textit{Pseudomonas putida}.

\begin{figure} \centering
\includegraphics[width=0.5\textwidth]{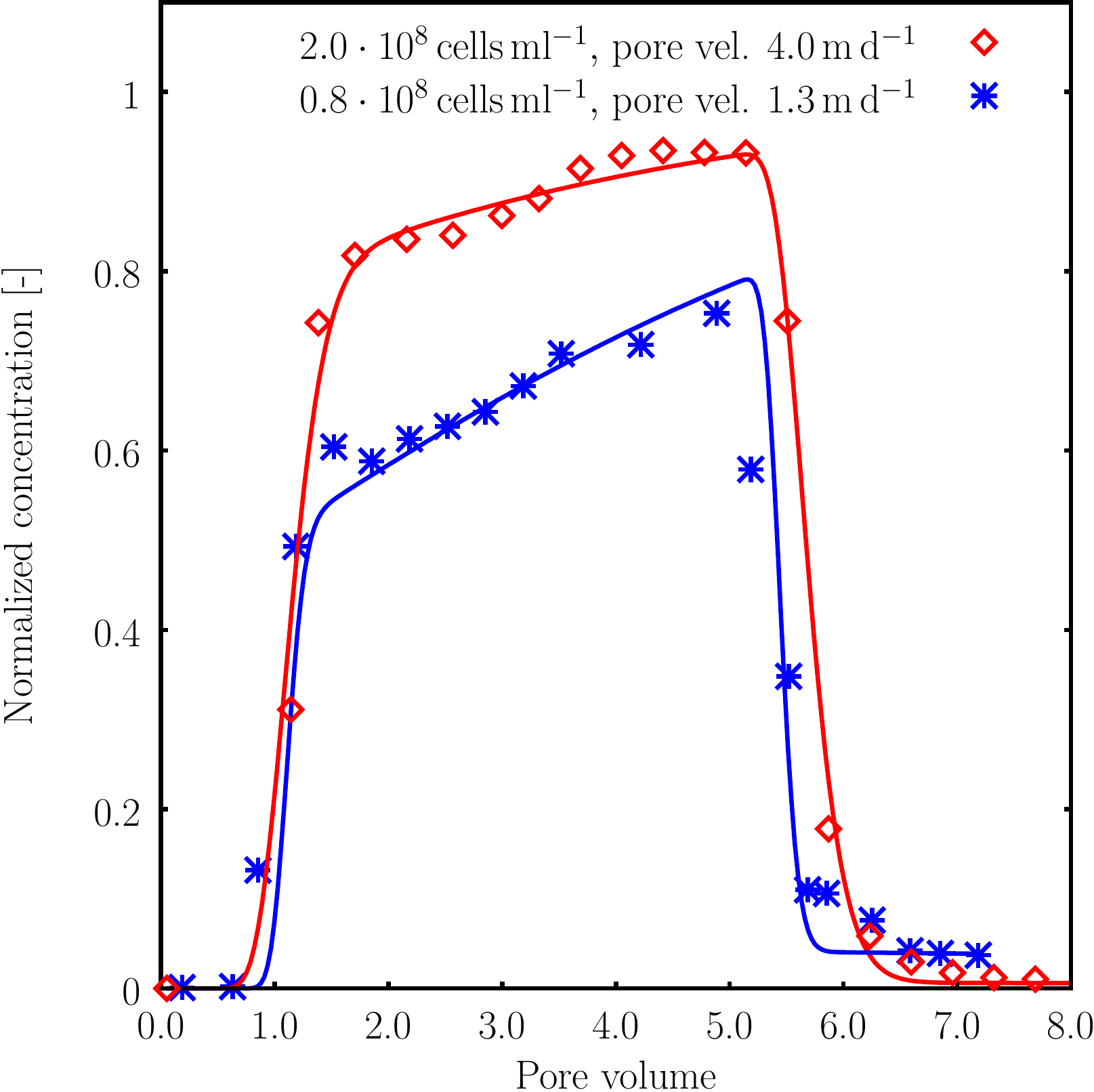}
 \caption{Flow through curves for \Ecoli{} HB101 K12 pGLO measured in
experiments under water-saturated conditions (points) and solution of
transport/adhesion problem with estimated parameters (lines).}
 \label{fig:adhesion}
\end{figure}

A comparison between measured flow through curves and the solution of
the transport/adhesion problem with given initial concentrations and
different pore velocities is given in Fig.~\ref{fig:adhesion}.

In the CFT the attachment rates are dependent on flow velocities, but
we have not observed any dependency on flow conditions. The influence
of bacterial growth stage on cell adhesion of \Ecoli{} has been
studied by \cite{Walker2004}. Cells in the stationary growth phase
were notably more adhesive than those in the exponential phase. In our
batch experiments, the age of \Ecoli{} cells did not have any
influence on cell adhesion kinetics.

\subsection{Growth parameters}

Growth parameters were estimated through inverse modelling based on
data from batch experiments. The estimated parameters are summarised
in Table~\ref{Tab:growth_parameters}.

\begin{table}[!h]
 \caption{Growth parameters}
 \centering
  \smallskip
  \small\addtolength{\tabcolsep}{5pt}
 \label{Tab:growth_parameters}
 \begin{tabular}{ l l} \toprule
  parameter& value \T \B
  \\ \midrule
  \mumaxa{} [\si{\per\hour}] & $0.324$ \T \\
  \mumaxan{} [\si{\per\hour}] & $ 0.255$ \T \\
  $\rd$ [\si{\per\hour}] & $\num{3.54e-3}$ \T \\
  \Bsa{} [-]  & $1.81$ \T \\
  \Bsan{} [-] & $3.07$ \T \\
  \Bo{} [-] & $ 0.019$ \T \\
  \Ysa{} [\si{\g \dw \per \g \consumed \doc}] &$0.95$ \T \\
  \Ysan{} [\si{\g \dw \per \g \consumed \doc}] & $0.163$ \T \\
  \Yo{} [\si{\g \dw \per \g \oxygen}] & $0.49$ \T \\
  \mo{} [\si{\cdm \per \hour \per \g \dw}] & $\num{0.003}$ \T \\ \bottomrule
\end{tabular}
\end{table}

The yield coefficients of \SI{0.95}{\g \dw \per \g} $\text{consumed
}\DOC$ and \num{0.49} \si{\g \dw \per \g \oxygen} estimated for the
\textit{E.~coli} cells are in accordance with the yield coefficients
reported by \cite{Reiling1985}, where also $\mu_{max,a}$ values
$0.39\text{-}\SI{1.39}{\per \hour}$ at high cultivation temperature of
$\SI{37}{\celsius}$ were determined. In \cite{Kovarova-Kovar1998}
values for $\mu_{max,an}$ between $0.19$ and $0.65$ in Chemostat and
batch cultures with glucose at $17-\SI{20}{\celsius}$ were listed.
Under anaerobic conditions bacteria use a mixed acid fermentation
(instead of respiration) to gain energy for growth \cite{Stokes1949,
Paege1961}, thus the growth is slower and the yield coefficient
decreases. According to our results, a yield of \SI{0.163}{\g \dw \per
\g \consumed \doc} was determined, which is close to the anaerobic
yield coefficient of \SI{0.18}{\g \dw \per \g \glucose} which
\cite{Ataai1985} found for some other strains of \textit{E.~coli}.

\subsection{Experimental setup}

To investigate growth and transport of \Ecoli{} under steady-state and
transient conditions, a flow through chamber (FTC) with inner dimensions of
$50\times40\times\SI{0.6}{\centi\meter}$ filled with sand up to
$\SI{30}{\centi \meter}$ was used. The whole system was kept under
constant temperature \temp{}. More detailed descriptions of the
laboratory experiment can be found in \cite{Jost2014}.

At the beginning of the experiment, the porous medium was completely
dry (previously heat sterilised). The suspension of \Ecoli{} cells and
LB medium was initially injected to the chamber through six ports in
the bottom of the chamber (at $5, 11.5, 21.5, 28, 39\text{ and }
\SI{44.5}{\centi\meter}$) with a flow-rate \SI{190}{\milli \litre \per
\hour}. After one hour, the water injection was stopped and the CF
at the height up to $\SI{25}{\centi \meter}$  was formed, see
saturation curve Fig. \ref{fig:NoflowExperimentComparison}. The
balance between water pressure and atmospheric pressure was at the
height of $\SI{6}{\centi \meter}$. The FTC was kept in standstill for
next five days, \Ecoli{} cells consumed all the bioconvertible \DOC{}
and synthesised green fluorescent protein (GFP).

After five days, sterile, oxygen saturated LB medium was injected from
the left side (inflow and outflow ports at height of $0.5, 2.0, 3.5
\text{ and } \SI{5.0}{\centi\meter}$) and the same amount of liquid
was extracted on the other side of the chamber by peristaltic
pumps. The total flow-rate in horizontal direction was $\SI{15}{\milli
\litre \per \hour}$ and took place for next six days.

The FTC was irradiated with UV light (\SI{365}{\nano
\metre}) and the mean fluorescence intensity, which was detected by a
camera system, was converted to cell densities, see
\cite{Jost2014}. The average dry weight $\weight$ of one cell of \Ecoli{}
was approx. \SI{5.e-13}{\gram}.

\subsection{Simulation setup}

Because the FTC is thin compared with the other
dimensions, for numerical simulations it can be simplified as a
two-dimensional domain $50\times\SI{30}{\centi\meter}$. The domain was
spatially discretized using $392 \times 256$ rectangular grid cells
for CCFV method, where the size of each element $1.3 \times
\SI{1.2}{\milli \meter}$ corresponds to the twofold diameter of
maximal grain size. This grid resolution is fine enough to eliminate
numerical error in spatial discretization, but the runtime of
numerical simulation is still acceptable.

In $p_c\text{-}p_l$-formulation of two-phase problem \eqref{Eq:PhaseBalance},
an initial positive water saturation is necessary (for $\Sl=0$ is
$p_c$ infinite). To obey this restriction, initial values of $p_c$ and
$p_l$ corresponds to $0.1\%$ of water content and atmospheric gas
pressure.

All domain boundaries excluding inlet/outlet ports are impermeable for
the liquid phase (zero Neumann boundary condition). The flux in ports
is identical to the experimental setup. For the gas phase all but
the top side is impermeable. The air pressure at the top of the domain
is fixed to the atmospheric pressure which is consistent to free
boundary.

The initial amount of water is free of \DOC{} and \Ecoli{} cells. The
relative initial and boundary oxygen concentration in air is
$20.95\%$. The ideal gas law yields absolute oxygen concentration of
$\SI{8.68}{\milli \mol \per \litre}$ gas. At $\SI{21}{\celsius}$ the
Henry's constant $\khcc$ is $\num{3.28e-2}$ \cite{dean1999} and the
equilibrium oxygen concentration $\coxys$ is $\SI{9.1}{\milli \gram
\per \litre}$~water.

The injected suspensions contain LB medium with $\SI{0.8}{\gram \per
\litre}$ of consumable \DOC{}. At the beginning of the experiment
(inflow phase), it additionally contains a negligible amount of
dissolved oxygen ($\SI{0.1}{\milli \gram \per \litre}$) and
$\SI{2e7}{\cells \per \milli \litre}$~water that is equivalent to a
biomass concentration of $\SI{1.e-2}{\milli \gram \per \milli
\litre}$. In the second part of the experiment the injected water does
not contain \Ecoli{} cells and dissolved oxygen concentration
corresponds to the equilibrium oxygen concentration $\coxys$.

For numerical simulations we assume, that \Ecoli{} cells are not able
to move without water flow (no molecular diffusion,
$D_{l,X}=\SI{0}{\meter \squared \per \second}$). Other diffusion
coefficients were taken from literature: $D_{l,S} =
\SI{1.9e-10}{\meter \squared \per \second}$ given by
\cite{Hendry2003}, $D_{l,O_2} = \SI{2.2e-9}{\meter \squared \per
\second}$ and $D_{g,O_2} = \SI{1.8e-5}{\meter \squared \per \second}$
as listed in \cite{Aachib2004}.

\subsection{Numerical simulation}

The numerical simulation was performed as described above. Our
interest is to have spatial and temporal distribution of investigated
species, primarily total cell concentration $\Xptotal=\Xtotal
\theta_l / \weight$~$ [\si{\cells \per \ccm \pvolume}]$, mobile cell
concentration $\Xp=\X \theta_l / \weight$~$
[\si{\cells \per \ccm \pvolume}]$, oxygen and \DOC{} concentrations in
order to understand the behaviour of flow and transport of \Ecoli{}
during the experiment. The CF volume denotes the total volume occupied
by the porous medium including the pore space.

\subsubsection*{Inflow and stagnancy}

The infiltration phase was stopped after $\SI{1}{\hour}$, but the complete
CF was established slower. After approx. $\SI{6}{\hour}$ no changes in
saturation distribution in both, in model as well as in experiment
were observed.

\begin{figure}[h] \centering
\includegraphics[width=0.5\textwidth]{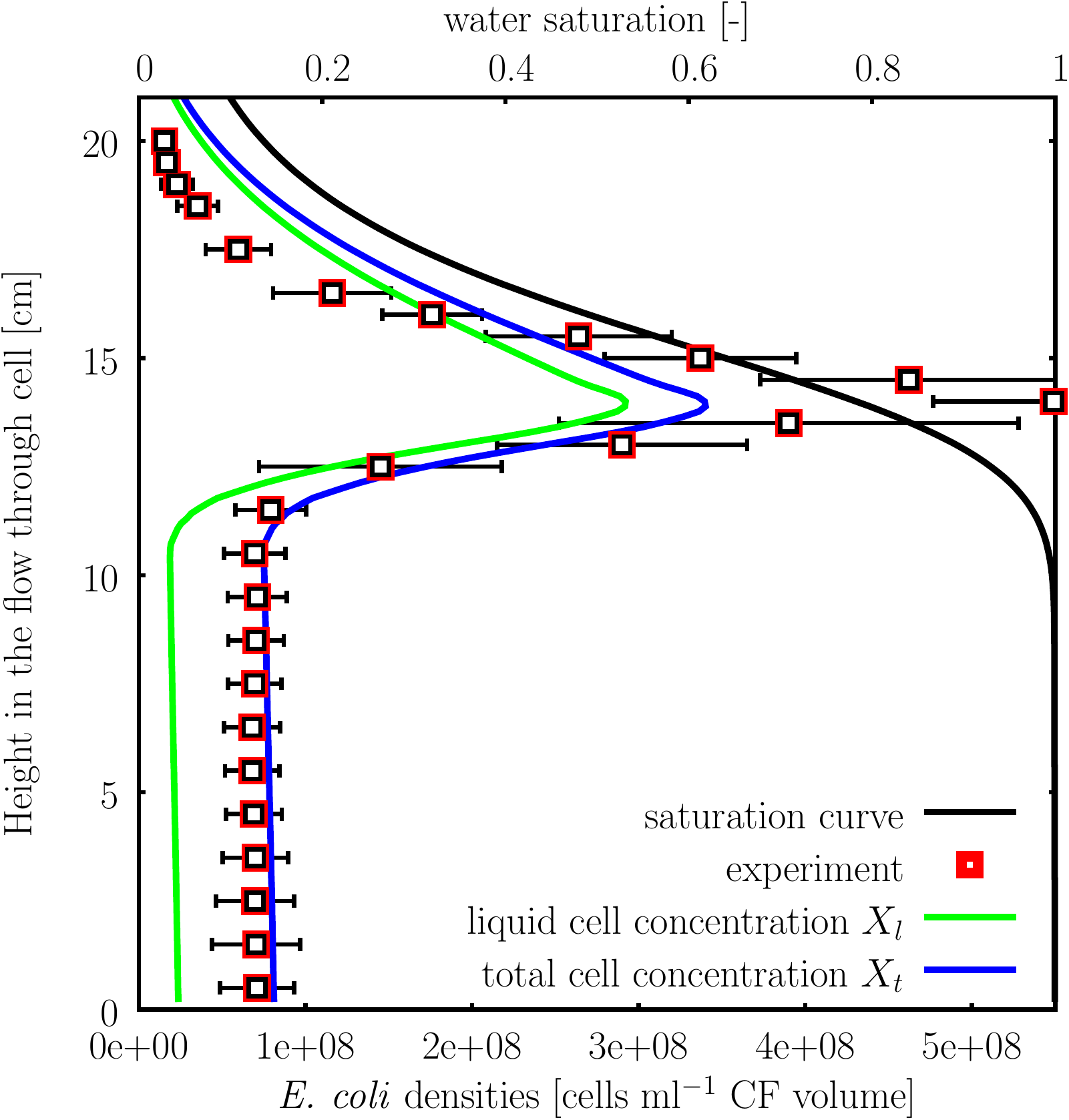}
  \caption{Vertical cut of the domain at $x=\SI{25}{\centi
      \meter}$ after $\SI{5}{d}$ of growth without new nutrient supply
    and without flow. Comparison between cell densities computed by
    numerical simulation with $\rp=\SI{0.9}{\milli \meter}$ and experimental data.}
 \label{fig:NoflowExperimentComparison}
\end{figure}

The concentration distributions in horizontal direction are roughly
symmetrical. The simulated cell densities on a vertical cut of the
two-dimensional domain at $x=\SI{25}{\cm}$ after $\SI{5}{\day}$ for
$\rp = \SI{0.9}{\milli \meter}$ are in
Fig.~\ref{fig:NoflowExperimentComparison}.

The lower part of the domain is up to \SI{11}{\cm} almost fully water
saturated. In this region the \Ecoli{} cells were growing without
oxygen supply anaerobically and the measured \Ecoli{} concentration is
\SI{0.7e8}{\cells \per \ccm \pvolume}. Because the attachment does not
depend on flow velocity and the time in stagnancy state is long, the
concentration of attached cells corresponds in the whole domain is
equal to $\Sl \Xsmax$.

Starting at the height of \SI{11}{\cm}, cells are growing under
anaerobic as well as aerobic conditions, because oxygen diffuses from
the air phase to liquid phase. The best growth activity and biomass
production was observed in the zone between \num{12} and \SI{16}{\cm},
where the water saturation is $0.6-0.8$. In this region, the cells
could optimally use the bioconvertible \DOC{}, which was initially
transported into the CF by infiltration and by capillary forces, and
the available dissolved oxygen.

The maximal cell density in laboratory experiment was higher than in
numerical simulation (\num{5.3e8} compared to \SI{3.6e8}{\cells \per
\ccm \pvolume}), but the uncertainty in computing cell densities from
their fluorescence intensities (FI) can be significant especially for
high FI intensities and the measured value is rather unrealistic. If
we take water saturation into account, the maximal cell concentration
in water at \SI{14}{\cm} is almost $11$ times higher compared to the
values in saturated region. Unfortunately, the yield coefficients for
aerobic growth measured in batch experiments are only $6$ time so
height to that for anaerobic growth.

In the upper part of the domain, the simulated
cell densities are proportional to the water saturation.
The higher
concentration than experimental data can be caused by overestimation
of oxygen phase exchange or by a slightly different shape of the
saturation curve in simulation and in laboratory experiment for lower
water saturation.

\begin{figure}[h] \centering
\includegraphics[width=0.5\textwidth]{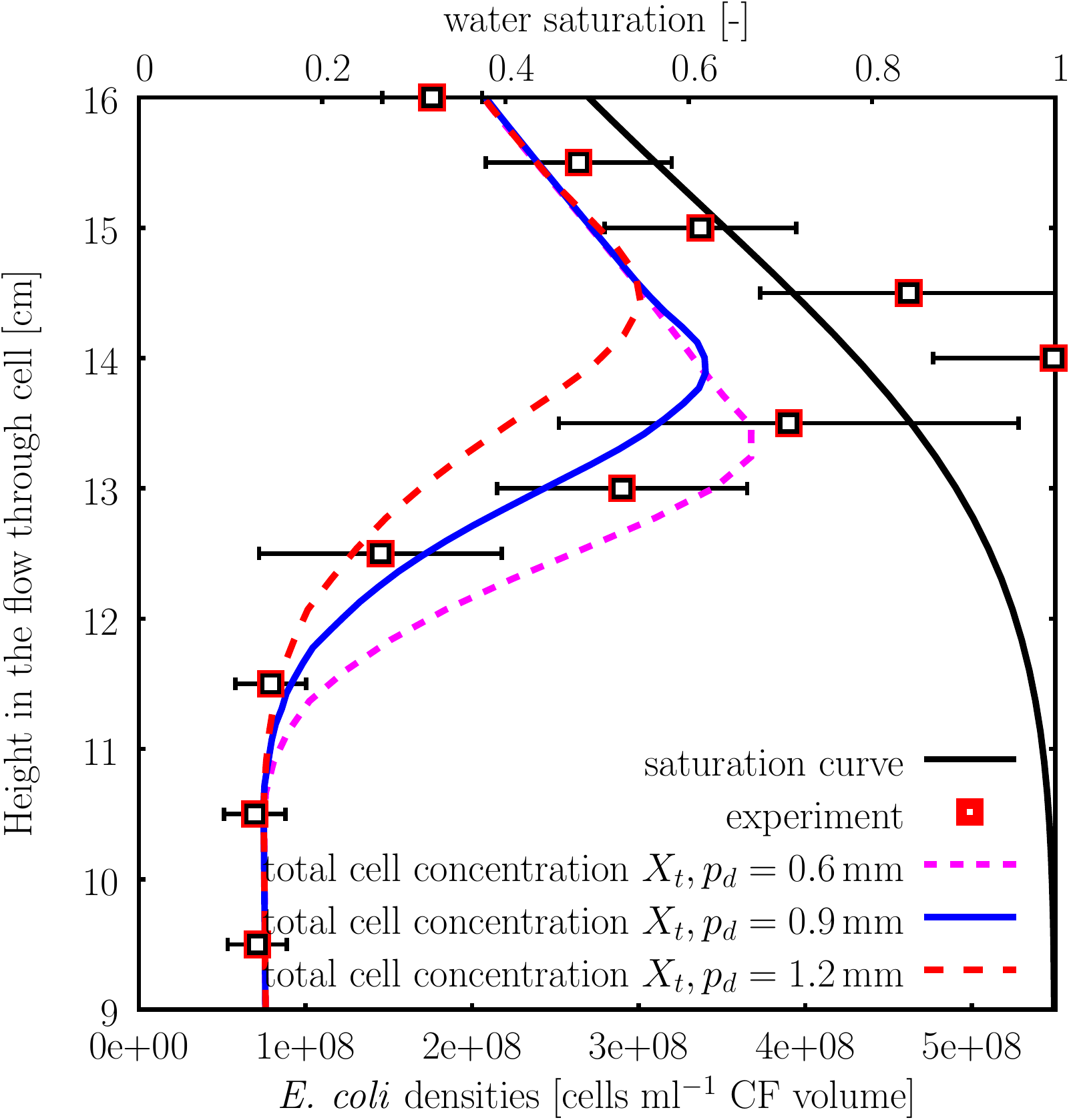}
  \caption{Vertical cut of the domain at $x=\SI{25}{\centi
      \meter}$ after $\SI{5}{d}$ of growth without new nutrient supply
    and without flow. Comparison between cell densities computed by
    numerical simulation  with different values of $\rp$  and experimental data.}
 \label{fig:NoflowExperimentComparison2}
\end{figure}

The simulated cell densities for different values of $\rp$, and thus
for different phase exchange rates, are compared in
Fig.~\ref{fig:NoflowExperimentComparison2}. The significant difference
was observed in regions with water saturation between $0.4$ and
$0.95$. For lower value of $\rp$, which corresponds to the size of the
bigger grains in the sand used, the phase exchange is faster in
regions with lower water saturation and model overpredicts the measured
cell densities. For large values of $\rp$ the maximal cell densities
decrease and the maximum is moved upwards in the domain. In following
numerical experiments, the value of $\SI{0.9}{\milli \meter}$ for
parameter $\rp$ was taken.

\subsubsection*{Horizontal flow}

After $\SI{5}{\day}$ in standstill, $\SI{15}{\milli \litre \per
\hour}$ of new oxygen saturated LB medium was injected to the
domain. In numerical simulations, tiny change of water saturation was
observed. On the inflow side of the domain, water saturation profile
rose by $\SI{4}{\milli \meter}$ and it decreased by the same value on
the outflow side of the chamber. This change in water saturation is too
small to be detected in laboratory experiment. After several hours,
the changes in water pore velocity were negligible and the flow field
and saturation distribution become stationary.

\begin{figure}[h] \centering
  \includegraphics[width=0.5\textwidth]{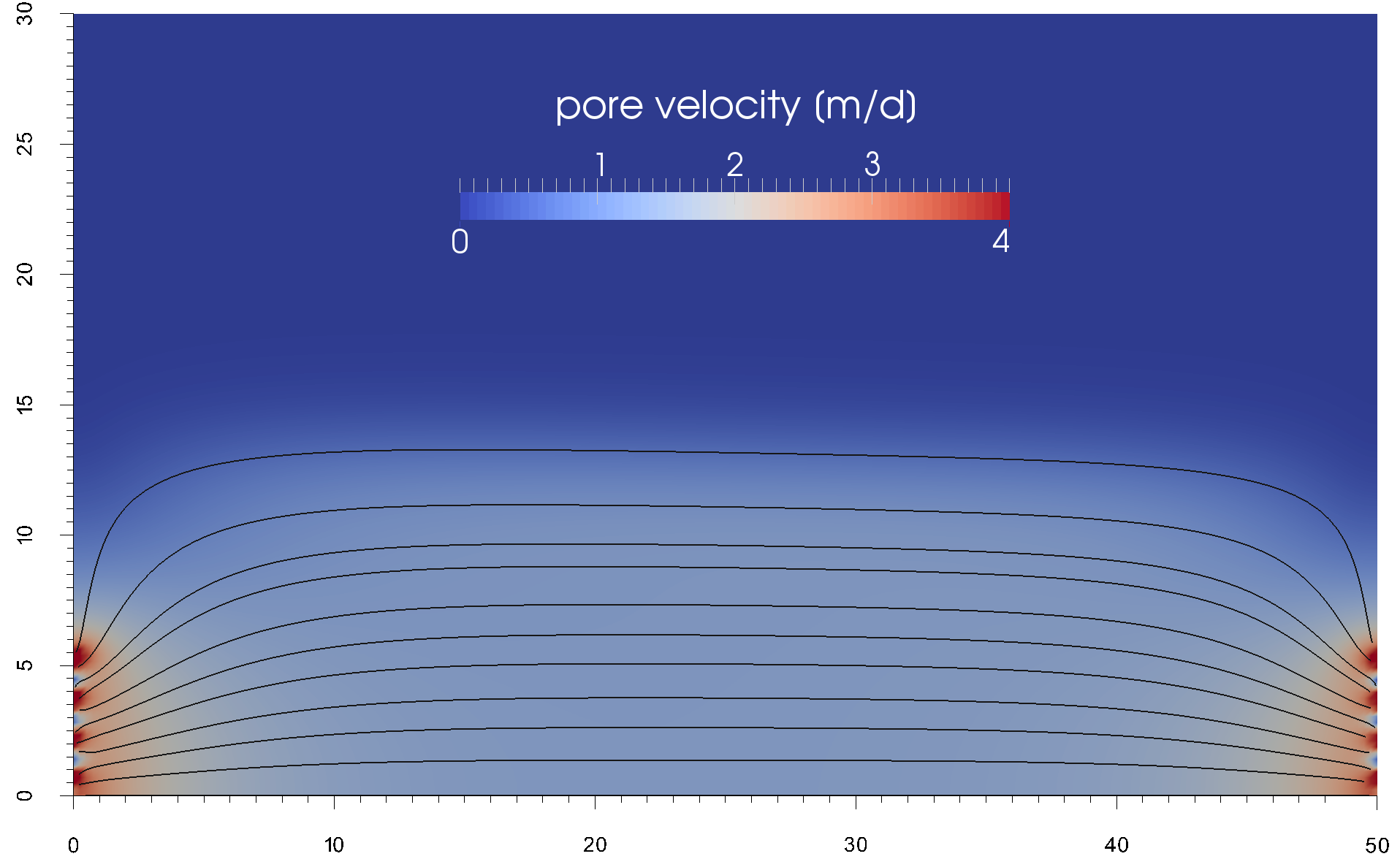}
  \caption{Magnitude of pore velocity flow field $[\si{\meter \per \day}]$
    (scaled to the maximal pore velocity $\SI{4}{\meter \per \day}$)
    and streamlines of water flow.}
 \label{fig:VelocityField}
\end{figure}

The computed water velocity flow field after reaching steady-state
flow conditions is shown in Fig.~\ref{fig:VelocityField}. In the
middle part of the domain up to $11-\SI{12}{\centi\meter}$, the water
flow is
quasi horizontal with pore velocity $\SI{1.3}{\meter \per \day}$. In
the unsaturated zone, the pore velocity is decreasing extensively with
water saturation reduction.

Close to inlets/outlets ports, the pore velocity reaches the magnitude
of $\SI{6.5}{\meter \per \day}$. Flow is not only horizontal, but it
follows the streamlines and new nutrients are transported to the upper
part of the domain and also cells in the unsaturated region will get
new \DOC{} supply.

Attachment kinetic causes cells attachment and all surface of sand
grains suitable for adhesion is filled up. New grown microorganisms
and detached cells become part of liquid phase and are transported
with water either to the unsaturated zone or are washed out of the
domain.

\begin{figure*}
\centering
  \includegraphics[width=\linewidth]{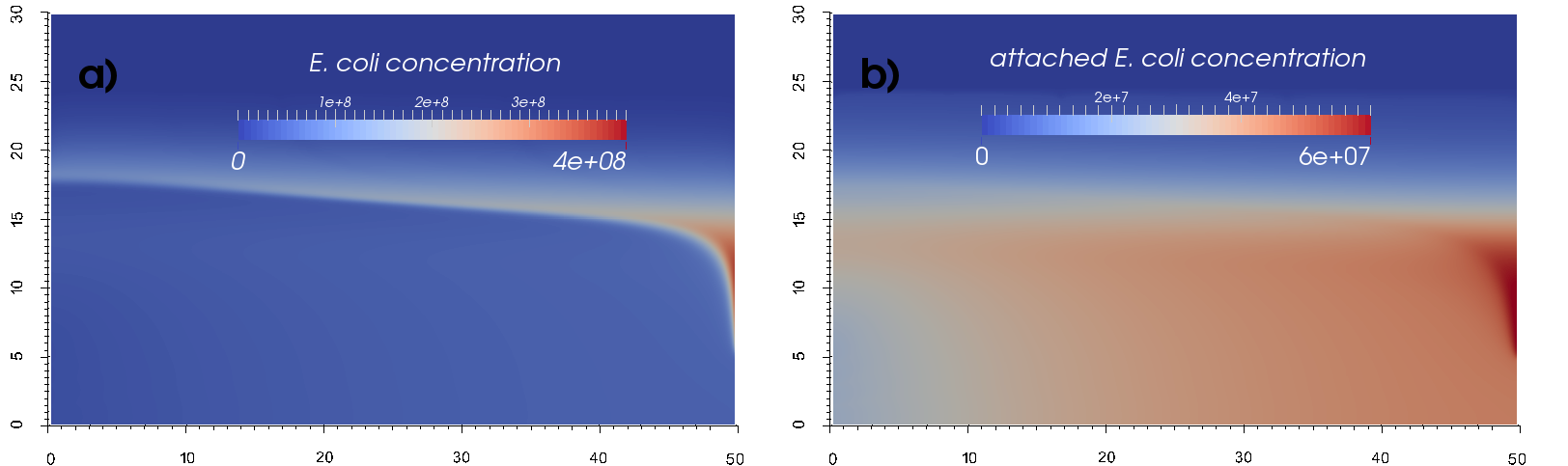}
  \label{fig:EcoliCells}
  \caption{Situation after 10 days of simulation without new \DOC{} supply, spatial
          distribution of a)~total \Ecoli{} concentration
    $\Xptotal$~$[\si{\cells \per \ml \pvolume}]$, b)~attached cells $[\si{\cells \per
      \ml \pvolume}]$.}
\label{fig:2DwithFlow2}
\end{figure*}

The result of numerical simulation without new \DOC{} supply, which is
a pure artificial case, is demonstrated in
Fig.~\ref{fig:2DwithFlow2}. The \Ecoli{} cells are highly washed away
of the FTC or they are moved to the the upper part of the domain, where the
fluid flow is insignificant. Even the cells originally attached to the
sand grains are slowly washed out, but to completely remove the
attached cells in water saturated regions, it would take much longer
time.

\begin{figure*}
\centering
  \includegraphics[width=\linewidth]{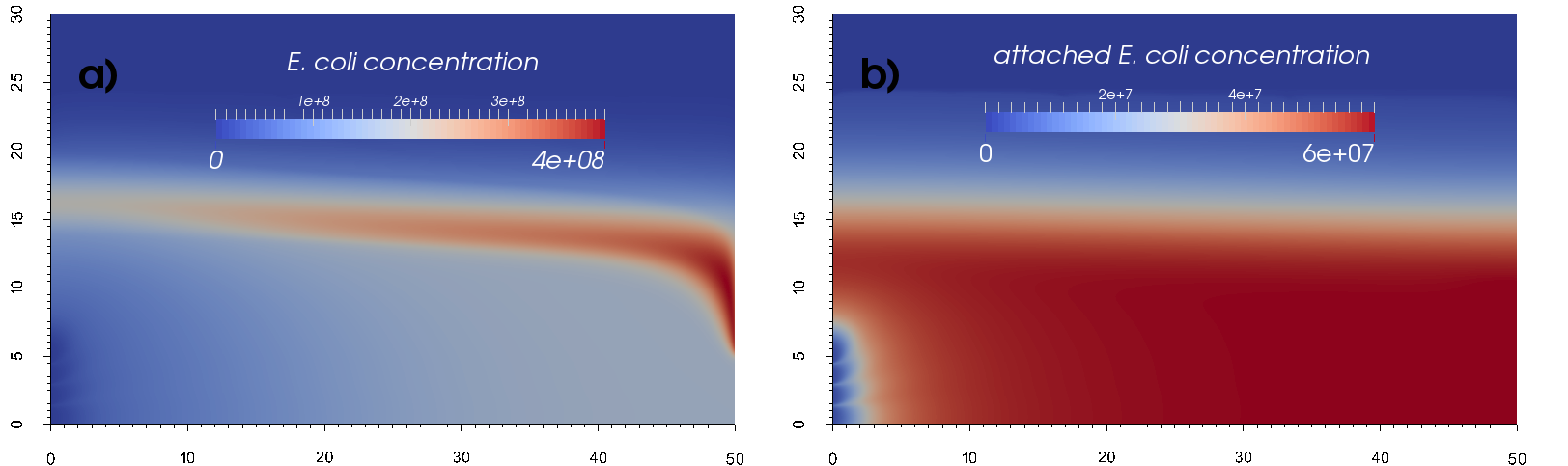}
  \label{fig:EcoliCells}
  \caption{Situation after 10 days of simulation with new \DOC{} supply, spatial
          distribution of a)~total \Ecoli{} concentration
    $\Xptotal$~$[\si{\cells \per \ml \pvolume}]$, b)~attached cells $[\si{\cells \per
      \ml \pvolume}]$.}
\label{fig:2DwithFlow}
\end{figure*}

The situation with new nutrients supply is completely different.  After
$\SI{3}{\day}$, the transport and growth of \Ecoli{} become
quasi-stationary and do not change considerably. The spatial
distribution after $\SI{10}{\day}$ (last $\SI{5}{\day}$ with
horizontal flow) is shown in Fig.~\ref{fig:2DwithFlow}.

On the inflow side of the chamber, the microorganisms are washed away
and even the number of attached cells is reduced. The area with high
cell concentration on the inflow side of the domain was shifted up by
$2-\SI{3}{\cm}$.

New carbon and oxygen sources are depleted during the transport
through the chamber. Oxygen in saturated and in unsaturated zone is
exhausted almost immediately if the microorganism concentration is
high. \DOC{} is transported with water flow and is consumed
continuously. Numerical simulation confirmed, that almost all
bioconvertible \DOC{} was consumed within transport through the
chamber.

\begin{figure}[t] \centering
\includegraphics[width=0.5\textwidth]{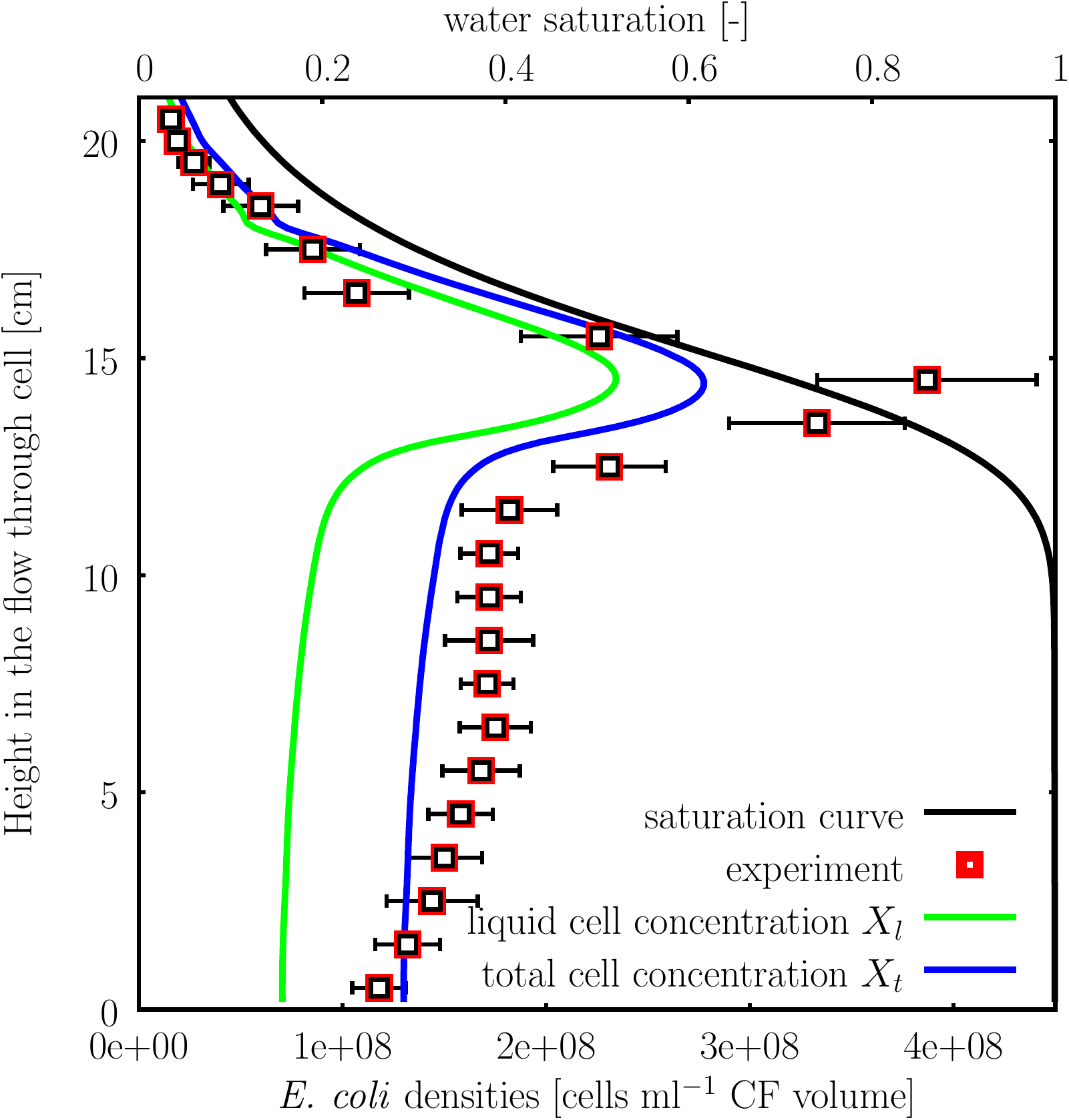}
  \caption{Vertical cut of the domain at $x=\SI{25}{\centi \meter}$
after $\SI{10}{d}$ of growth, last $\SI{5}{d}$ with new nutrient
supply and horizontal flow. Comparison between cell densities computed
by numerical simulation and experimental data.}
  \label{fig:FlowExperimentComparison}
\end{figure}

The comparison between experimental and simulated cell concentrations
on a vertical cut at $x=\SI{25}{\cm}$ is in
Fig.~\ref{fig:FlowExperimentComparison}. With numerical simulation it
is possible to predict the total cell concentration well. In the upper
part of the domain with very low saturation ($\Sl<0.2$) \Ecoli{} cells
did not get any new nutrients from flowing water and were
starving. The highest microbial activity is again in the transition
zone between $12 \text{ and } \SI{16}{\cm}$, but the maximal cell
concentration is lower than without flow
(Fig.~\ref{fig:NoflowExperimentComparison}). The vertical flow in the
saturated region causes cell attachment and the total cell
concentration $\Xptotal$ is higher than the mobile cell concentration
$\Xp$; in laboratory experiment even higher cell numbers were
measured.

Numerical model predicts an accumulation of \Ecoli{} cells in the
right part of the domain above the outflow ports. From this region the
cells are washed away slowly. This effect was also observed in tracer
experiments with fluorescein, but not observed in the experiment with
\Ecoli{} cells, see \cite{Jost2014b}.

In addition, the production of GFP is dependent on oxygen
concentration \cite{Jost2014}, but many cells were grown without
oxygen supply which can limit the GFP production and in situ cell
detection can be difficult in the right part of the domain. These
contradiction between numerical simulation and experimental data
should be taken into account and be investigate in the design of
future laboratory experiments.

\section{Discussion and summary}

The primary objective of this study was to provide a macroscopic
approach for modelling microbial growth and transport in saturated and
unsaturated porous media. In pursuing this objective we have
emphasised two areas of advancement. The first is development of
numerical simulator for two-phase multicomponent reactive flow, and
the second is model optimisation and comparison between numerical
simulation and laboratory experiment data. The simulations were
performed without any additional calibration of the estimated and of
the used parameters.

\subsection*{Model formulation and numerical discretization}

The model includes transient two phase movement in
saturated-unsaturated porous media, multicomponent solute transport
and diffusion, phase exchange, as well as a chemical model that
considers chemical reactions, microbial growth and cell attachment.

The operator splitting concept allows us to use discretization schemes
that are adapted in an optimal way to the individual situation
dependent on flow conditions. Since reaction only takes place when
components mix, the numerical diffusion in flow direction can
overpredict pore-scale mixing, and thereby overpredict homogeneous
chemical reaction \cite{Gramling2002}.

To prevent numerical dispersion and overestimation of the reactions,
higher order discretization schemes are used. Using explicit temporal
discretization, the numerical diffusion in advective-dominated cases
is reduced significantly, but the time step is restricted by the $CFL$
condition.

If the flow is slow (diffusion dominates) or if the flow is
quasi-stationary and takes place for a longer time (several pore
volumes), the DSA together with implicit time discretization can be an
alternative. Although the resulting system of nonlinear algebraic
equations can be large, the time step is not restricted. In practise,
the time step for the DSA is restricted by the convergence of Newton's
method.

As the geometry of the experimental setup is rectangular, simple
structured grids are sufficient. A robust, flexible, accurate, local
mass conservative and efficient space discretization scheme on such
grids is the CCFV. However, on complicated geometries and unstructured
grids other discretization schemes like discontinuous Galerkin method
can be used \cite{Bastian2014}.

\subsection*{Microbiological growth}

In order to predict microbial processes in the field or laboratory
situation, we need to understand the dominate processes in the
particular setting and obtain estimates of kinetic parameters.

Microbial growth and decay, cell attachment and detachment, phase
exchange and transport of nutrients - these are the most important
processes affecting behaviour of \Ecoli{} in the capillary
fringe. Model describing \Ecoli{} growth under aerobic and anaerobic
conditions and its combination based on observations in batch
experiments was furthermore extended by transport and adhesion
processes. Because the advection is in our application high and a
significant part of microorganisms remains in the liquid phase, we do
not need to introduce a maximal microbial capacity in growth model as
in \cite{Schaefer1998}.

By simulating \Ecoli{} growth and transport and subsequent consumption
of \DOC{} and oxygen, it was possible to reproduce the observed changes of biomass like in the laboratory experiment. The microbial
activity was observed and modelled in the whole domain, particularly
in the transition zone. In this region, the bacteria will get new \DOC{}
supply with water flow and sufficient amount of oxygen from the
atmosphere.

\subsection*{Model limitations}

In our macroscopic modelling approach,
we neglected or simplified some processes which may affect growth and
transport of \Ecoli{} and other soil microorganisms.

For adhesion kinetic we use a model based on colloid filtration
theory, which is not always valid under repulsive conditions
\cite{Tufenkji2007}. Although we assumed the same consumption kinetic
in both phases, attached bacteria may consume less nutrients than
cells in liquid phase. In addition, the attachment mechanism can be
different in water fully saturated and unsaturated regions.

More understanding is also required on how soluble organics interact
with the surfaces of bacteria and affect their retention to soil
surfaces \cite{Unc2004}. In \cite{Schafer1998,Jiang2007} the
accumulation of bacteria at the air-water interface was also
considered and it was modelled as an irreversible adsorption. We did
not deem it necessary in this study, because no attachment and
detachment from the air-water interface was observed in laboratory
experiments.

We also neglected the capability of microorganisms to move in response
to a chemical gradient. Both random motility and chemotaxis can have
an influence to transport for subsurface organisms
\cite{Ginn2002}. Motility and chemotaxis can play a role e.g. in
movement of microbes into contaminated areas of low permeability,
where advective transport is minimal compared with adjacent preferred
flow paths \cite{Mills1997}.

The growth and transport of microorganisms can also have an influence
on hydraulic properties of porous media
\cite{Yarwood2006}. High accumulation of microbial cells can change the
hydraulic properties of porous media \cite{Taylor1990}, create
preferential flow paths and change flow and transport direction, which
is not yet included in the model. The microbially generated gases and
entrapment or microbially induced changes in the chemical properties
of the liquid or solid phases of the media may also play a role in
permeability reduction \cite{Rockhold2002}.

The models describing the formation of microcolonies and/or biofilms
require microscale data that are difficult to measure accurately
\cite{Clement1998}. Microbial colonisation may cause apparent drying
within the colonised zone, with localised decreases in saturation,
giving rise to partial diversion of flow around the colonised zone
\cite{Yarwood2006}.

\section{Conclusion}

The developed model provides a temporally and spatially resolved
quantification of microbial activity and nutrients in unsaturated and
saturated porous medium. The individual submodels were optimised and
calibrated based on experimental data and used for the simulation of
growth and transport of \Ecoli{} in a flow through cell with
conditions representing the capillary fringe. The results of numerical
simulation can reproduce laboratory experiments very effectively. The model
can be further extended and applied to optimise the design of new
laboratory and field experiments.

{\small \noindent
\subsection*{Acknowledgements}
This study was funded by DFG (German Research Foundation) through
the Research Group FOR 831 ``Dynamic Capillary Fringes: A
Multidisciplinary Approach'' (Project Ga 546/5-2 and Ba
1498/7-2).
}

{\small
\bibliographystyle{spmpsci}
\bibliography{references}
}
\end{document}